\documentclass[aps,prd,showpacs,nofootinbib,preprint]{revtex4}

\usepackage{amsmath}
\usepackage{amssymb}
\usepackage{graphicx}

\begin{document}

\title{Galactic dark matter as a bulk effect on the brane}

\author{C. G. B\"ohmer}
\email{christian.boehmer@port.ac.uk}
\affiliation{Institute of Cosmology \& Gravitation,
             University of Portsmouth, Portsmouth PO1 2EG, UK}

\author{T. Harko}
\email{harko@hkucc.hku.hk}
\affiliation{Department of Physics and Center for Theoretical
             and Computational Physics, The University of Hong Kong,
             Pok Fu Lam Road, Hong Kong}

\date{\today}

\begin{abstract}
The behavior of the angular velocity of a test particle moving in
a stable circular orbit in the vacuum on the brane is considered.
In the brane world scenario, the four dimensional effective
Einstein equation acquire extra terms, called dark radiation and
dark pressure, respectively, which arise from the embedding of the
3-brane in the bulk. A large number of independent observations
have shown that the rotational velocities of test particles
gravitating around galaxies tend, as a function of the distance
from the galactic center, toward constant values. By assuming a
constant tangential velocity, the general solution of the vacuum
gravitational field equations on the brane can be obtained in an
exact analytic form. This allows us to obtain the explicit form of
the projections of the bulk Weyl tensor on the brane, and the
equation of state of the dark pressure as a function of the dark
radiation. The physical and geometrical quantities are expressed
in terms of observable/measurable parameters, like the tangential
velocity, the baryonic mass and the radius of the galaxy. We also
analyze the dynamics of test particles by using methods from the
qualitative analysis of dynamical systems, by assuming a simple
linear equation of state for the dark pressure. The obtained
results provide a theoretical framework for the observational
testing at the extra-galactic scale of the predictions of the
brane world models.
\end{abstract}


\maketitle

\section{Introduction}

The problem of the dark matter is a long standing problem in
modern astrophysics. Two important observational issues, the
behavior of the galactic rotation curves and the mass discrepancy
in clusters of galaxies led to the necessity of considering the
existence of dark matter at a galactic and extra-galactic scale.

The rotation curves of spiral galaxies~\cite{Bi87} are one of the
best evidences showing the problems Newtonian gravity and/or
standard general relativity have to face on the
galactic/intergalactic scale. In these galaxies neutral hydrogen
clouds are observed at large distances from the center, much
beyond the extent of the luminous matter. Since these clouds move in
circular orbits with velocity $v_{tg}(r)$, the orbits are
maintained by the balance between the centrifugal acceleration
$v_{tg}^2/r$ and the gravitational attraction $GM(r)/r^2$ of
the total mass $M(r)$ contained within the orbit. This allows the
expression of the mass profile of the galaxy in the form
$M(r)=rv_{tg}^2/G$.

Observations show that the rotational velocities increase near the
center of the galaxy, in agreement with the theory, but then remain
nearly constant at a value of $v_{tg\infty }\sim 200-300$ km/s~\cite{Bi87},
which leads to a mass profile $M(r)=rv_{tg\infty }^2/G$.
Consequently, the mass within a distance $r$ from the center of
the galaxy increases linearly with $r$, even at large distances
where very little luminous matter has been detected.

The second astrophysical evidence for dark matter comes from the
study of the clusters of galaxies. The total mass of a cluster can
be estimated in two ways. Knowing the motions of its member
galaxies, the virial theorem gives one estimate, $M_{VT}$, say.
The second is obtained by separately estimating the mass of each
individual members, and summing these masses, to give a total
baryonic mass $M_{B}$. Almost without exception it is found that
$M_{VT}$ is considerably greater than $M_{B}$, $M_{VT}>M_{B}$,
typical values of $M_{VT}/M_{B}$ being about 20-30~\cite{Bi87}.

This behavior of the galactic rotation curves and of the virial
mass of galaxy clusters is usually explained by postulating the
existence of some dark (invisible) matter, distributed in a
spherical halo around the galaxies. The dark matter is assumed to
be a cold, pressure-less medium. There are many possible
candidates for dark matter, the most popular ones being the weakly
interacting massive particles (WIMP) (for a review of the particle
physics aspects of the dark matter see~\cite{OvWe04}). Their
interaction cross sections with normal baryonic matter, while
extremely small, are expected to be non-zero, and we may expect to
detect them directly.

It has also been suggested that the dark matter in the Universe
might be composed of super-heavy particles, with mass $\geq
10^{10}$ GeV. But observational results show that the dark matter
can be composed of super-heavy particles only if these interact
weakly with normal matter, or if their mass is above $10^{15}$
GeV~\cite{AlBa03}. Scalar fields or other long range coherent
fields coupled to gravity have also intensively been used to model
galactic dark matter~\cite{scal}.

A general analysis of the possibility of an alternative
four-dimensional gravity theory explaining the dynamics of
galactic systems without dark matter was performed by Zhytnikov
and Nester \cite{Ne94}. From very general assumptions about the
structure of a relativistic gravity theory (the theory is metric,
and invariant under general coordinates transformation, has a good
linear approximation, it does not possess any unusual gauge
freedom and it is not a higher derivative gravity) a general
expression for the metric to order $(v/c)^2$ has been derived.
This allows to compare the predictions of the theory with various
experimental data: the Newtonian limit, light deflection and
retardation, rotation of galaxies and gravitational lensing. The
general conclusion of this study is that the possibility for any
four-dimensional gravity theory to explain the behavior of
galaxies without dark matter is rather improbable.

However, up to now no non-gravitational evidence for dark matter
has been found, and no direct evidence or annihilation radiation
from it has been observed yet.

Therefore, it seems that the possibility that Einstein's (and the Newtonian)
theory of gravity breaks down at the scale of galaxies cannot be excluded
\textit{a priori}. Several theoretical models, based on a modification of
Newton's law or of general relativity, have been proposed to explain the
behavior of the galactic rotation curves~\cite{expl}.

The idea of embedding our Universe in a higher dimensional
space has attracted a considerable interest recently, due to the
proposal by Randall and Sundrum~\cite{RS99a} that our
four-dimensional (4d) spacetime is a three-brane, embedded in a
5d spacetime (the bulk).  According to the
brane world scenario, the physical fields (electromagnetic,
Yang-Mills etc.) in our 4d Universe are confined to
the three brane. Only gravity can freely propagate in both the
brane and bulk spacetimes, with the gravitational self-couplings
not significantly modified. Even if the fifth dimension is
uncompactified, standard 4d gravity is reproduced on the brane.
Hence this model allows the presence of large, or even infinite
non-compact extra dimensions. Our brane is identified to a domain
wall in a 5d anti-de Sitter spacetime. For a review of the dynamics
and geometry of brane Universes, see e.g.~\cite{GeMa01}.

Due to the correction terms coming from the extra dimensions,
significant deviations from the standard Einstein theory occur in
brane world models at very high energies~\cite{SMS00}. Gravity is
largely modified at the electro-weak scale of about 1~TeV. The
cosmological and astrophysical implications of the brane world
theories have been extensively investigated in the physical
literature~\cite{all2,sol}.

 The static vacuum gravitational field equations on the brane depend on the generally
unknown Weyl stresses, which can be expressed in terms of two
functions, called the dark radiation $U$ and the dark pressure $P$
terms (the projections of the Weyl curvature of the bulk,
generating non-local brane stresses)~\cite{Da00,Mar04,GeMa01}.

Several classes of spherically symmetric solutions of the static
gravitational field equations in the vacuum on the brane have been
obtained in~\cite{Ha03,Ma04,Ha05}. As a possible physical
application of these solutions the behavior of the angular velocity $%
v_{tg}$ of the test particles in stable circular orbits has been
considered~\cite{Ma04,Ha05}. The general form of the solution,
together with two constants of integration, uniquely determines
the rotational velocity of the particle. In the limit of large
radial distances, and for a particular set of values of the
integration constants the angular velocity tends to a constant
value. This behavior is typical for massive particles (hydrogen
clouds) outside galaxies~\cite{Bi87}, and is usually explained by
postulating the existence of the dark matter.

Thus, the rotational galactic curves can be naturally explained in
brane world models, without introducing any additional hypothesis.
The galaxy is embedded in a modified, spherically symmetric
geometry, generated by the non-zero contribution of the Weyl
tensor from the bulk. The extra terms, which can be described in
terms of the dark radiation term $U$ and the dark pressure term
$P$, act as a ``matter'' distribution outside the galaxy. The
particles moving in this geometry feel the gravitational effects
of $U$, which can be expressed in terms of an equivalent mass (the
dark mass) $M_U$. The dark mass is linearly increasing with the
distance, and proportional to the baryonic mass of the galaxy,
$M_{U}(r)\approx M_{B}(r/r_{0})$~\cite{Ma04}.

The exact galactic metric, the dark radiation and the dark pressure in
the flat rotation curves region in the brane world scenario has been
obtained in~\cite{Ha05}.

Similar interpretations of the dark matter as bulk effects have
been also considered in~\cite{Pal}, where it was also shown that
it is possible to model the X-ray profiles of clusters of
galaxies, without the need for dark matter.

In the brane world scenario, the fundamental scale of gravity is
not the Planck scale, but another scale which may be at the TeV
level. The gravitons propagating through the bulk space give rise
to a Kaluza-Klein tower of massive gravitons on the brane. These
gravitons couple to the energy-momentum term of the standard model
fields, and could be produced under the appropriate circumstances
as real or virtual particles.

Another important effect that is expected in the brane world
models is the presence of brane fluctuations, since rigid objects
do not exist in the relativistic theory. It was proposed that, in
the context of brane world scenarios with low tension $\tau=f^4$,
massive brane fluctuations are natural dark matter candidates,
called branons \cite{CeDoMa03}. The different possibilities for
branons as dark matter candidates, the parameter region in which
branons behave as collisionless thermal relics (WIMPs), either
cold or hot (warm) together with less standard scenarios in which
they are strongly self interacting or produced non-thermally, and
the possibilities of obtaining branons in hadron accelerators have
been considered in \cite{CeDoMa2}. Another possibility to test
these models is through the annihilation of branons into photon or
electron-positron pairs in the galactic halos, which could
possibly be detected by gamma-ray or anti-matter detectors.

A possible brane world model, which tries to give a justification
for the hypothesis of the scalar field origin for the dark matter
was proposed by \cite{Matetal}. The model contains two branes, on
one of the branes lives the matter of the standard model of
particles but on the other one, only spin-0 fundamental
interactions are present. In the model the spin-0 fields are the
dark matter. The scalar field contains an effective pressure,
which avoids the collapse of the field fluctuations, implying that
scalar field dominated objects, like galaxies, may contain a flat
density profile in the center.

Gravitational lensing and the study of the light deflection by
black holes and galaxies is an important physical effect that
could provide specific signatures for testing the brane world
models (for a review of the gravitational lensing by brane world
black holes see \cite{Majum05}). Observables related to the
relativistic images of strong field gravitational lensing could in
principle be used to distinguish between different brane world
black hole metrics in future observations.  The strong field limit
approach was used in \cite{Wh05} to investigate the gravitational
lensing properties of brane world black holes, and the lensing
observables for some candidate brane world black hole metrics have
been compared with those for the standard Schwarzschild case.

Brane World black holes could have significantly different lensing
observational signatures as compared to the Schwarzschild black
holes. The bending angle predicted by the brane world models is
much larger than that predicted by standard general relativistic
and dark matter models \cite{Majum05,Wh05}. The deflection of
photons and the bending angle of light in the constant tangential
velocity region on the brane was considered in \cite{Pal,Ha05}.
The bending angle predicted by the brane world models is much
larger than that predicted by standard general relativistic and
dark matter models. Therefore, the study of the gravitational
lensing could discriminate between the different dynamical laws
proposed to model the motion of particles at the galactic level
and the standard dark matter models.

Generally, the vacuum field equations on the brane can be reduced
to a system of two ordinary differential equations, which describe
all the geometric properties of the vacuum as functions of the
dark pressure and dark radiation terms~\cite{Ha03}. In order to
close the system of vacuum field equations on the brane a
functional relation between these two quantities is necessary.

Hence, a first possible approach to the study of the vacuum brane
consists in adopting an explicit equation of state for the dark
pressure as a function of the dark radiation. A second method
consist in closing the field equations on the brane by imposing
the condition of the constancy of the rotational velocity curves
for particle in stable orbits.

It is the purpose of the present paper to consider the general
behavior of the vacuum gravitational field equations in the brane
world model in the region of constant tangential rotational
velocity of test particles in stable circular orbits. Physically,
this situation is characteristic for particles gravitating in
circular orbits around the galactic center~\cite{Bi87}. As a first
step in our study we derive, under the assumption of spherical
symmetry, the basic equations describing the structure of the
vacuum on the brane, and the equation giving the behavior of the
tangential velocity of the test particles as functions of the dark
radiation and of the dark pressure.

To obtain the tangential velocity of test particles we analyze the
motion in an effective potential, also containing the
effects of the bulk. By assuming that the brane is a fixed point
of the bulk, we derive the equation giving the tangential velocity
as a function of the $00$ component of the metric tensor only. In
the region of constant tangential velocities, the general
solutions of the gravitational equations can be obtained in an
exact analytic form.

This allows us to obtain the explicit form of the projections of
the bulk Weyl tensor on the brane (the dark radiation and the dark
pressure), as well as the components of the metric tensor. The
mass associated to the dark radiation (the dark mass) has a
similar behavior as the dark matter at the galactic scale -- it is
a linearly increasing function of the distance, and it is
proportional to the square of the tangential velocity. This result
strongly supports the interpretation of the usual ``dark matter''
as a bulk effect. All the physical and geometrical quantities in
our model are expressed in terms of observable/measurable
parameters, like the tangential velocity, the baryonic mass and
the radius of the galaxy. Some astrophysical applications of our
model, including the possibilities of its observational testing,
are also briefly considered.

Since generally the structure equations of the vacuum on the brane
cannot be solved exactly, we shall analyze the dynamics of test
particles by using methods from the qualitative analysis of
dynamical systems. In particular, we will investigate the general
behavior of the tangential velocity on the brane by assuming a
simple linear equation of state for the dark pressure. In this
case generally we cannot reproduce the observed behavior of the
test particles.

The present paper is organized as follows. The field equations for
the vacuum and the tangential velocity of a test particle on the brane 
are written down in Sections~\ref{seci} and~\ref{secii}. The general 
solution of the field equations in the constant tangential velocity region 
is obtained in Section~\ref{seciii}. The qualitative analysis of the 
field equations for a linear equation of state for the dark pressure is
performed in Section~\ref{seciv}. We discuss and conclude our results in
Section~\ref{secv}.

\section{The field equations and the tangential velocity of test particles
for static, spherically symmetric vacuum branes}
\label{seci}

\subsection{The field equations in the brane world models}

We start by considering a five dimensional (5d) spacetime (the
bulk), with a single four-dimensional (4d) brane, on which matter
is confined, only gravity can probe the extra dimensions. The 4d
brane world $({}^{(4)}M,g_{\mu\nu})$ is
located at a hypersurface $\left(B\left(X^{A}\right) =0\right)$
in the 5d bulk spacetime $({}^{(5)}M,g_{AB})$, of which
coordinates are described by $X^{A},A=0,1,\ldots,4$. The induced 4d
coordinates on the brane are $x^{\mu},\mu =0,1,2,3$.

The action of the system is given by~\cite{SMS00}
\begin{equation}
S=S_{bulk}+S_{brane},  \label{bulk}
\end{equation}
where
\begin{equation}
S_{bulk}=\int_{{}^{(5)}M}\sqrt{-{}^{(5)}g}
\left[\frac{1}{2k_{5}^{2}}{}^{(5)}R+{}^{(5)}L_{m}+\Lambda_{5}\right] d^{5}X,
\end{equation}
and
\begin{equation}
S_{brane}=\int_{{}^{(4)}M}\sqrt{-{}^{(5)}g}
\left[\frac{1}{k_{5}^{2}}K^{\pm}+L_{brane}\left( g_{\alpha \beta },\psi \right)+
\lambda_{b}\right] d^{4}x,
\end{equation}
where $k_{5}^{2}=8\pi G_{5}$ is the 5d gravitational constant,
${}^{(5)}R$ and ${}^{(5)}L_{m}$ are the 5d scalar curvature and
the matter Lagrangian in the bulk, $L_{brane}\left( g_{\alpha
\beta },\psi\right)$ is the 4d Lagrangian, which is given by a
generic functional of the brane metric $g_{\alpha \beta }$ and of
the matter fields $\psi $, $K^{\pm }$ is the trace of the
extrinsic curvature on either side of the brane, and $\Lambda
_{5}$ and $\lambda _b$ (the constant brane tension) are the
negative vacuum energy densities in the bulk and on the brane,
respectively.

The Einstein field equations in the bulk are given by~\cite{SMS00}
\begin{equation}
{}^{(5)}G_{IJ}=k_{5}^{2} {}^{(5)}T_{IJ},\qquad {}^{(5)}T_{IJ}=-\Lambda _{5}
{}^{(5)}g_{IJ}+\delta(B)\left[-\lambda_{b} {}^{(5)}g_{IJ}+T_{IJ}\right] ,
\end{equation}
where
\begin{equation}
{}^{(5)}T_{IJ}\equiv - 2\frac{\delta {}^{(5)}L_{m}}{\delta {}^{(5)}g^{IJ}}
+{}^{(5)}g_{IJ} {}^{(5)}L_{m},
\end{equation}
is the energy-momentum tensor of bulk matter fields, while $T_{\mu \nu }$ is
the energy-momentum tensor localized on the brane and which is defined by
\begin{equation}
T_{\mu \nu }\equiv -2\frac{\delta L_{brane}}{\delta g^{\mu \nu }}+g_{\mu \nu}\, L_{brane}.
\end{equation}

The delta function $\delta \left( B\right) $ denotes the
localization of brane contribution. In the 5d spacetime a brane is
a fixed point of the $Z_{2}$ symmetry. The basic equations on the
brane are obtained by projections onto the brane world. The
induced 4d metric is $g_{IJ}={}^{(5)}g_{IJ}-n_{I}n_{J}$, where
$n_{I}$ is the space-like unit vector field normal to the brane
hypersurface ${}^{(4)}M$. In the following we assume ${}^{(5)}L_{m}=0$.

Assuming a metric of the form $ds^{2}=(n_{I}n_{J}+g_{IJ})dx^{I}dx^{J}$, with
$n_{I}dx^{I}=d\chi $ the unit normal to the $\chi = {\rm constant}$ hypersurfaces
and $g_{IJ}$ the induced metric on $\chi = {\rm constant}$ hypersurfaces, the
effective 4d gravitational equation on the brane (the Gauss equation),
takes the form~\cite{SMS00}:
\begin{equation}
G_{\mu \nu }=-\Lambda g_{\mu \nu }+k_{4}^{2}T_{\mu \nu} +
k_{5}^{4}S_{\mu \nu}-E_{\mu \nu },  \label{Ein}
\end{equation}
where $S_{\mu \nu }$ is the local quadratic energy-momentum correction
\begin{equation}
S_{\mu \nu }=\frac{1}{12}TT_{\mu \nu }-\frac{1}{4}T_{\mu}{}^{\alpha }
T_{\nu \alpha }+\frac{1}{24}g_{\mu \nu }\left( 3T^{\alpha \beta }
T_{\alpha \beta}-T^{2}\right) ,
\end{equation}
and $E_{\mu \nu }$ is the non-local effect from the free bulk gravitational
field, the transmitted projection of the bulk Weyl tensor $C_{IAJB}$, $%
E_{IJ}=C_{IAJB}n^{A}n^{B}$, with the property $E_{IJ}\rightarrow E_{\mu \nu
}\delta _{I}^{\mu }\delta _{J}^{\nu }\quad $as$\quad \chi \rightarrow 0$. We
have also denoted $k_{4}^{2}=8\pi G$, with $G$ the usual 4d gravitational constant.

The 4d cosmological constant, $\Lambda $, and the 4d coupling
constant, $k_{4}$, are related by $\Lambda =k_{5}^{2}(\Lambda
_{5}+k_{5}^{2}\lambda _{b}^{2}/6)/2$ and
$k_{4}^{2}=k_{5}^{4}\lambda _{b}/6$, respectively. In the limit
$\lambda_{b}^{-1}\rightarrow 0$ we recover standard general
relativity \cite{SMS00}.

The Einstein equation in the bulk and the Codazzi equation also imply the
conservation of the energy-momentum tensor of the matter on the brane, $%
D_{\nu }T_{\mu }{}^{\nu }=0$, where $D_{\nu }$ denotes the brane
covariant derivative. Moreover, from the contracted Bianchi
identities on the brane it follows that the projected Weyl tensor
obeys the constraint $D_{\nu }E_{\mu }{}^{\nu
}=k_{5}^{4}D_{\nu }S_{\mu }{}^{\nu }$.

The symmetry properties of $E_{\mu \nu }$ imply that in general we can
decompose it irreducibly with respect to a chosen $4$-velocity
field $u^{\mu}$ as~\cite{Mar04}
\begin{equation}
E_{\mu \nu }=-k^{4}\left[ U\left( u_{\mu }u_{\nu} +\frac{1}{3}h_{\mu \nu
}\right) +P_{\mu \nu }+2Q_{(\mu }u_{\nu)}\right] ,  \label{WT}
\end{equation}
where $k=k_{5}/k_{4}$, $h_{\mu \nu }=g_{\mu \nu }+u_{\mu }u_{\nu }$ projects
orthogonal to $u^{\mu }$, the ``dark radiation'' term $U=-k^{-4}E_{\mu \nu
}u^{\mu }u^{\nu }$ is a scalar, $Q_{\mu }=k^{-4}h_{\mu }^{\alpha }E_{\alpha
\beta }u^{\beta}$ is a spatial vector and $P_{\mu \nu }=-k^{-4}\left[ h_{(\mu }\text{ }%
^{\alpha }h_{\nu )}\text{ }^{\beta }-\frac{1}{3}h_{\mu \nu }h^{\alpha \beta }%
\right] E_{\alpha \beta }$ is a spatial, symmetric and trace-free
tensor.

In the case of the vacuum state we have $\rho =p=0$, $T_{\mu \nu }\equiv 0$,
and consequently $S_{\mu \nu }=0$. Therefore the field equation describing
a static brane takes the form
\begin{equation}
R_{\mu \nu }=-E_{\mu \nu }+\Lambda g_{\mu \nu },
\end{equation}
with the trace $R$ of the Ricci tensor $R_{\mu \nu }$ satisfying the
condition $R=R_{\mu }^{\mu }=4\Lambda $.

In the vacuum case $E_{\mu \nu }$ satisfies the constraint $D_{\nu }E_{\mu
}{}^{\nu }=0$. In an inertial frame at any point on the brane we have $%
u^{\mu }=\delta _{0}^{\mu}$ and $h_{\mu\nu}={\rm diag}(0,1,1,1)$.
In a static vacuum $Q_{\mu}=0$ and the constraint for $E_{\mu\nu}$
takes the form~\cite{GeMa01}
\begin{equation}
\frac{1}{3}D_{\mu }U+\frac{4}{3}UA_{\mu }+D^{\nu }P_{\mu \nu }+A^{\nu}
P_{\mu \nu }=0,
\end{equation}
where $A_{\mu }=u^{\nu }D_{\nu }u_{\mu }$ is the
4-acceleration. In the static spherically symmetric case we may chose $%
A_{\mu }=A(r)r_{\mu }$ and $P_{\mu \nu }=P(r)\left( r_{\mu }r_{\nu }-\frac{1%
}{3}h_{\mu \nu }\right) $, where $A(r)$ and $P(r)$ (the ``dark
pressure'' although the name dark anisotropic stress might be more
appropriate) are some scalar functions of the radial distance~$r$,
and~$r_{\mu }$ is a unit radial vector~\cite{Da00}.

\subsection{The motion of particles in stable circular orbits on the brane}

In brane world models test particles are confined to the brane.
Mathematically, this means that the equations governing the motion
are the standard 4d geodesic equations \cite{Seahra:2002xe,Mar04}. However, the
bulk has an effect on the motion of the test particles on the
brane via the metric. Since the projected Weyl tensor effectively
serves as an additional matter source, the metric is affected by
these bulk effects, and so are the geodesic equations. This is in
contrast to Kaluza-Klein theories, where matter travels on 5d
geodesics.

In order to obtain results which are relevant to the galactic dynamics,
in the following we will restrict our study to the static and spherically
symmetric metric given by
\begin{equation}
ds^{2}=-e^{\nu(r)}dt^{2}+e^{\lambda(r)}dr^{2}+
r^{2}d\Omega^2,
\label{metr1}
\end{equation}
where $d\Omega^2 = d\theta^{2}+\sin^{2}\theta d\phi^{2}$.

The Lagrangian $\mathcal{L}$ for a massive test particle
traveling on the brane reads
\begin{equation}
\mathcal{L}=\frac{1}{2}\left(-e^{\nu}\dot{t}^{2}+e^{\lambda}\dot{r}^{2}+
r^{2}\dot{\Omega}^2\right),
\label{lag}
\end{equation}
where the dot means differentiation with respect to the affine parameter.

Since the metric tensor coefficients do not explicitly depend on
$t$ and $\Omega$, the Lagrangian~(\ref{lag}) yields the following
conserved quantities (generalized momenta):
\begin{equation}
-e^{\nu (r) }\dot{t}=E,\qquad
r^2 \dot{\Omega} = L,
\label{cons}
\end{equation}
where $E$ is related to the total energy of the particle and $L$ to the total angular
momentum. With the use of conserved quantities we obtain from Eq.~(\ref{lag}) the
geodesic equation for massive particles in the form
\begin{equation}
e^{\nu +\lambda }\dot{r}^2 + e^{\nu}\left(1+\frac{L^2}{r^2}\right) = E^{2},
\label{geod1}
\end{equation}

The second term of the left-hand side can, in some cases, be
interpreted as an effective potential. For instance, for the
Schwarzschild space-time, where $e^{\nu+\lambda}=1$, the kinetic
term is position independent. In that case the notion of an
effective potential is appropriate. In other cases, even one can
still compute the turning points of the kinetic term, however, the
effective potential interpretation is lost.

For the case of the motion of particles in circular and stable
orbits the `potential' must satisfy the following
conditions: a) $\dot{r}=0$ (circular motion) b) $\partial
V_{eff}/\partial r$ $=0$ (extreme motion) and c) $\partial
^{2}V_{eff}/\partial r$ $^{2}\left.\right|_{extr}>0$ (stable
orbit), respectively. Conditions a) and b) immediately give the
conserved quantities as
\begin{equation}
E^{2}=e^{\nu }\left(1+\frac{L^{2}}{r^{2}}\right),
\label{cons1}
\end{equation}
and
\begin{equation}
\frac{L^2}{r^{2}} = \frac{r\nu'}{2} e^{-\nu} E^2,
\label{cons2}
\end{equation}
respectively. Equivalently, these two equations can be rewritten as
\begin{equation}
E^2 = \frac{e^\nu}{1-r\nu'/2}, \qquad
L^2 = \frac{r^3 \nu'/2}{1-r\nu'/2}.
\end{equation}

We define the tangential velocity $v_{tg}$ of a test particle on
the brane, as measured in terms of the proper time, that is, by an
observer located at the given point, as~\cite{LaLi}
\begin{equation}
\label{vtgbr}
v_{tg}^{2}=e^{-\nu }r^{2}\left(\frac{d\Omega}{dt}\right)^2 = \frac{e^{\nu}}{E^2}\frac{L^2}{r^2}.
\end{equation}

By using the constants of motion, and eliminating the quantity
$L$, we obtain the expression of the tangential velocity of a test
particle in a stable circular orbit~\cite{Matos} on the brane as
\begin{equation}
\label{vtg}
v_{tg}^{2}=\frac{r\nu'}{2}.
\end{equation}

Let us emphasize again that the function $\nu'$ is obtained by
solving the field equations containing the bulk effects as
additional matter terms; we consider this in Section~\ref{secii}.

\subsection{The gravitational field equations for a static
spherically symmetric brane}

With the metric given by~(\ref{metr1}) the gravitational field equations and
the effective energy-momentum tensor conservation equation in the vacuum take
the form~\cite{Ha03,Ma04}
\begin{equation}
-e^{-\lambda }\left( \frac{1}{r^{2}}-\frac{\lambda ^{\prime }}{r}\right) +%
\frac{1}{r^{2}}=\frac{48\pi G}{k^{4}\lambda _{b}}U+\Lambda ,
\label{f1}
\end{equation}
\begin{equation}
e^{-\lambda }\left( \frac{\nu ^{\prime }}{r}+\frac{1}{r^{2}}\right) -\frac{1%
}{r^{2}}=\frac{16\pi G}{k^{4}\lambda _{b}}\left( U+2P\right) -\Lambda ,
\label{f2}
\end{equation}
\begin{equation}
e^{-\lambda }\frac{1}{2}\left( \nu ^{\prime \prime }+\frac{\nu ^{\prime 2}}{2%
}+\frac{\nu ^{\prime }-\lambda ^{\prime }}{r}-\frac{\nu ^{\prime }\lambda
^{\prime }}{2}\right) =\frac{16\pi G}{k^{4}\lambda _{b}}\left( U-P\right)
-\Lambda ,  \label{f3}
\end{equation}
\begin{equation}
\nu ^{\prime }=-\frac{U^{\prime }+2P^{\prime }}{2U+P}-\frac{6P}{r\left(
2U+P\right) },  \label{f4}
\end{equation}
where $^{\prime }=d/dr$. In the following we denote
\begin{equation}
\alpha =\frac{16\pi G}{k^{4}\lambda _{b}}.
\end{equation}

As for the motion of the test particle on the brane we assume that
they follow stable circular orbits, with tangential velocities
given by Eq. (\ref{vtg}). Thus, the rotational velocity of the
test body is determined by the metric coefficient $\exp \left( \nu
\right) $ only.

The field equations~(\ref{f1})--(\ref{f2}) yield the following effective
energy density, radial and orthogonal pressure
\begin{eqnarray}
\rho ^{\mathrm{eff}} &=&3\alpha U+\Lambda ,  \label{eq1} \\
P^{\mathrm{eff}} &=&\alpha U+2\alpha P-\Lambda ,  \label{eq2} \\
P_{\perp }^{\mathrm{eff}} &=&\alpha U-\alpha P-\Lambda ,  \label{eq3}
\end{eqnarray}
which by rewriting become
\begin{equation}
\rho ^{\mathrm{eff}}-P^{\mathrm{eff}}-2P_{\perp }^{\mathrm{eff}}=4\Lambda =%
\mathrm{const.}
\end{equation}

This is expected for the `radiation' like source, the projection
of the bulk Weyl tensor, which is trace-less $E_\mu^\mu = 0$.

\section{Structure equations of the vacuum in the brane world models}
\label{secii}

Eq.~(\ref{f1}) can immediately be integrated to give
\begin{equation}
e^{-\lambda }=1-\frac{C_{1}}{r}-\frac{GM_{U}\left( r\right) }{r}-\frac{%
\Lambda }{3}r^{2},  \label{m1}
\end{equation}
where $C_{1}$ is an arbitrary constant of integration, and we denoted
\begin{equation}
GM_{U}\left( r\right) =3\alpha \int_{0}^{r}U(r)r^{2}dr.
\end{equation}

The function $M_U$ is the gravitational mass corresponding to the
dark radiation term (the dark mass). For $U=0$ the metric
coefficient given by Eq.~(\ref{m1}) must tend to the standard
general relativistic Schwarzschild metric coefficient, which gives
$C_{1}=2GM$, where $M = {\rm constant}$ is the baryonic (usual)
mass of the gravitating system.

By substituting $\nu'$ given by Eq.~(\ref{f4}) into Eq.~(\ref{f2})
and with the use of Eq.~(\ref{m1}) we obtain the following system
of differential equations satisfied by the dark radiation term
$U$, the dark pressure $P$ and the dark mass $M_U$, describing the
vacuum gravitational field, exterior to a massive body, in the
brane world model \cite{Ha03}:
\begin{equation}
\frac{dU}{dr}=-\frac{\left( 2U+P\right) \left[ 2GM+GM_{U}+\alpha
\left(
U+2P\right) r^{3}\right] -\frac{2}{3}\Lambda r^{2}}{r^{2}\left( 1-\frac{2GM}{%
r}-\frac{M_{U}}{r}-\frac{\Lambda }{3}r^{2}\right) }-2\frac{dP}{dr}-\frac{6P}{%
r},  \label{e1}
\end{equation}
\begin{equation}
\frac{dM_{U}}{dr}=\frac{3\alpha }{G}r^{2}U.  \label{e2}
\end{equation}

The tangential velocity can be written as
\begin{equation}
v_{tg}^{2}=\frac{1}{2}\frac{2GM+GM_{U}+\alpha \left( U+2P\right) r^{3}-\frac{2%
}{3}\Lambda r^{2}}{r\left( 1-\frac{2GM}{r}-\frac{GM_{U}}{r}-\frac{\Lambda }{3}%
r^{2}\right) }.  \label{vtg2}
\end{equation}

To close the system a supplementary functional relation between
one of the unknowns $U$, $P$, $M_{U}$ and $v_{tg}$ is needed. Once
this relation is known, Eqs.~(\ref{e1})--(\ref{vtg}) give a full
description of the geometrical properties and of the motion of the
particles on the brane.

The system of equations~(\ref{e1}) and~(\ref{e2}) can be transformed to an
autonomous system of differential equations by means of the transformations
\begin{equation}
q=\frac{2GM}{r}+\frac{GM_{U}}{r}+\frac{\Lambda }{3}r^{2},\qquad
\mu=3\alpha r^{2}U+3r^{2}\Lambda ,\nonumber
\end{equation}
\begin{equation}
p=3\alpha r^{2}P-3r^{2}\Lambda ,\qquad \theta =\ln r. \label{trans}
\end{equation}
We shall call $\mu $ and $p$ the ``reduced'' dark radiation and pressure,
respectively.

With the use of the new variables given by Eqs.~(\ref{trans}), Eqs.~(\ref{e1})
and~(\ref{e2}) become
\begin{equation}
\frac{dq}{d\theta }=\mu -q,  \label{aut1}
\end{equation}
\begin{equation}
\frac{d\mu }{d\theta }=-\frac{\left( 2\mu +p\right) \left[ q+\frac{1}{3}%
\left( \mu +2p\right) \right] }{1-q}-2\frac{dp}{d\theta }+2\mu -2p.
\label{aut2}
\end{equation}

Eqs.~(\ref{e1}) and~(\ref{e2}), or, equivalently,~(\ref{aut1})
and~(\ref {aut2}) may be called the structure equations of the
vacuum on the brane. In order to close this system an ``equation
of state'', relating the reduced dark radiation and the dark
pressure terms is needed. Generally, this equation of state is
given in the form $P=P(U)$.

In the new variables the tangential velocity of a particle in a stable
circular orbit on the brane is given by
\begin{equation}  \label{vtgx}
v_{tg}^2=\frac{1}{2}\frac{q+\frac{1}{3}\left( \mu +2p\right)}{1-q}.
\end{equation}
By using the expression of the tangential velocity, Eq.~(\ref{aut2}) can be
rewritten as
\begin{equation}  \label{muvtg}
\frac{d}{d\theta }\left(\mu +2p\right)=-2\left( 2\mu +p\right)
v_{tg}^{2}+2\mu -2p.
\end{equation}

Eq.~(\ref{muvtg}) allows the easy check of the physical
consistency of some simple equations of state for the dark
pressure. The equation of state $\mu +2p=0$ immediately gives
$v_{tg}^2=1$, implying that all test particle in stable circular
motion on the brane move at the speed of light, a fact that is
contradicted by the observations at the galactic scale. The
equation of state $2\mu +p=0$ gives $\mu =\mu _0/r^2$, where $\mu
_0$ = constant is an arbitrary integration constant, $U=\mu
_0/3\alpha r^4$ and $GM_U=-\mu _0/3r^3$, respectively. In the
limit of large $r$, the tangential velocity $v_{tg}^2$ tends to
zero, $v_{tg}\rightarrow 0$. Therefore, this model seems also to
be ruled out by observations. The case $\mu =p$ gives $\mu
\left(\theta \right)=\mu _0\exp\left[-2\int{v_{tg}^2\left(\theta
\right)}d\theta \right]$, $\mu _0=\mathrm{constant}$, and
$q(\theta )=\left(2v_{tg}^2-\mu \right)/\left(1+2v_{tg}^2\right)$.

The dark radiation and the dark pressure can be obtained as a
functions of the tangential velocity in a closed analytical form
in the important case of a linear equation of state of the form
\begin{equation}
p=\left( \Gamma -2\right) \mu +\beta ,
\end{equation}
with $\Gamma $ and $\beta $ arbitrary constants. Then the reduced
dark radiation can be obtained as
\begin{multline}
\mu \left( \theta \right)=\theta ^{2\left( 3-\Gamma \right)
/\left( 2\Gamma -3\right) }\exp \left[ -\frac{2\Gamma }{2\Gamma
-3}\int v_{tg}^{2}\left( \theta \right) d\theta \right] \\\times
\left\{ C_0-\frac{3\beta }{2\Gamma -3}\int \left[
1+v_{tg}^{2}\left( \theta \right) \right] \theta ^{-2\left(
3-\Gamma \right) /\left( 2\Gamma -3\right) }\exp \left[
\frac{2\Gamma }{2\Gamma -3}\int v_{tg}^{2}\left( \theta \right)
d\theta \right] \right\},
\end{multline}
where $C_0$ is an arbitrary integration constant. Hence, if the
velocity profile of a test particle in stable circular motion is
known, one can obtain all the relevant physical parameters for a
static spherically symmetric system on the brane.

\section{Dark radiation, dark pressure and the galactic metric
in the constant tangential velocity region}
\label{seciii}

The galactic rotation curves provide the most direct method of
analyzing the gravitational field inside a spiral galaxy. The
rotation curves have been determined for a great number of spiral
galaxies. They are obtained by measuring the frequency shifts $z$
of the light emitted from stars and from the 21cm radiation
emission from the neutral gas clouds. Usually, the astronomers
report the resulting $z$ in terms of a velocity field $v_{tg}$.
The observations show that at distances large enough from the
galactic center
\begin{equation}
v_{tg}\approx 200-300\text{km/s}=\mathrm{constant.}
\end{equation}
This behavior has been observed for a large number of galaxies~\cite{Bi87}.

In the following we use this observational constraint to
reconstruct the metric of a galaxy on the brane. The constancy of
$v_{tg}$ allows us to express the function $q$ from
Eq.~(\ref{vtg}) as
\begin{equation}
q=\frac{2v_{tg}^{2}}{1+2v_{tg}^{2}}-\frac{1}{3\left( 1+2v_{tg}^{2}\right) }%
\left( \mu +2p\right),  
\label{q}
\end{equation}
giving immediately
\begin{equation}
\frac{dq}{d\theta }=-\frac{1}{3\left( 1+2v_{tg}^{2}\right) }\left( \frac{%
d\mu }{d\theta }+2\frac{dp}{d\theta }\right).  
\label{int}
\end{equation}

From Eq.~(\ref{muvtg}) we obtain
\begin{equation}
\frac{d\mu }{d\theta }+2\frac{dp}{d\theta }=-2\left(
1-2v_{tg}^{2}\right) \mu -2\left( 1+v_{tg}^{2}\right) p,
\end{equation}
which can be used to simplify Eq.~(\ref{int}) into
\begin{equation}  
\label{autn}
\frac{dq}{d\theta }=-\frac{2}{3\left( 1+2v_{tg}^{2}\right) }\left[ \left(
1-2v_{tg}^{2}\right) \mu -\left( 1+v_{tg}^{2}\right) p\right].
\end{equation}

By eliminating $dq/d\theta $ between Eqs.~(\ref{aut1}), and
(\ref{aut2}) and by using the expression of $q$ given by
Eq.~(\ref{q}), we obtain the equation of state for the dark
pressure in the constant tangential velocity region as
\begin{equation}
p=\frac{3+v_{tg}^{2}}{v_{tg}^{2}}\mu -3.
\label{eos}
\end{equation}
This shows that, as a function of the ``reduced'' dark radiation,
in the coordinate $\theta $, the ``reduced'' dark pressure obeys a
generalized linear equation of state.

By using the generalized linear equation of state, Eq.~(\ref{aut2}) can be
written as
\begin{equation}
\frac{d\mu }{d\theta }=-n\mu +m,  \label{eqfin}
\end{equation}
where
\begin{equation}
n=\frac{2v_{tg}^{2}}{3\left( 2+v_{tg}^{2}\right) }\left[ \frac{3+v_{tg}^{2}%
}{v_{tg}^{2}}+3v_{tg}^{2}+2\right] ,
\end{equation}
and
\begin{equation}
m=\frac{2\left( 1+v_{tg}^{2}\right) v_{tg}^{2}}{2+v_{tg}^{2}}.
\end{equation}
Eq.~(\ref{eqfin}) can be immediately integrated to give
\begin{equation}
\mu \left( \theta \right) =Ce^{-n\theta }+\frac{m}{n},
\end{equation}
with $C$ an arbitrary constant of integration.

In the standard radial coordinate $r$ the ``reduced'' dark radiation energy
density is given by
\begin{equation}
\mu (r)=\frac{C}{r^{n}}+\frac{\left( 1+v_{tg}^{2}\right) v_{tg}^{2}}{%
1+v_{tg}^{2}+v_{tg}^{4}},
\end{equation}
while the ``reduced'' dark pressure is
\begin{equation}
p(r)=\frac{3+v_{tg}^{2}}{v_{tg}^{2}}\frac{C}{r^{n}}-\frac{2v_{tg}^{2}}{%
1+v_{tg}^{2}}.
\end{equation}

From these equations we find the variation of the dark radiation
and dark pressure in the constant tangential velocity region as
\begin{equation}
U(r)=\frac{1}{3\alpha }\left[ \frac{C}{r^{n+2}}+\frac{\left(1+v_{tg}^{2}%
\right) v_{tg}^{2}} {1+v_{tg}^{2}+v_{tg}^{4}}\frac{1}{r^{2}} - 3\Lambda %
\right] ,
\end{equation}
and
\begin{equation}
P(r)=\frac{1}{3\alpha }\left[ \frac{3+v_{tg}^{2}}{v_{tg}^{2}}\frac{C}{r^{n+2}%
} -\frac{2v_{tg}^{2}}{1+v_{tg}^{2}}\frac{1}{r^{2}} + 3\Lambda\right] ,
\end{equation}
respectively.

The metric tensor component $\exp(\nu)$ can be found from the expression of
the tangential velocity as
\begin{equation}
e^{\nu }=C_{2}r^{2v_{tg}^{2}},
\end{equation}
with $C_2$ an arbitrary constant of integration, while $\exp(\lambda)$ is
given by
\begin{equation}
e^{-\lambda } = 1-\frac{2GM}{r}-\frac{C}{\left( 1-n\right)r^{n}} -\frac{%
\left( 1+v_{tg}^{2}\right)v_{tg}^{2}}{1+v_{tg}^{2}+v_{tg}^{4}} -\frac{\Lambda%
}{3} r^2.
\end{equation}

The function $M_{U}(r)$, which gives the mass associated with the dark
radiation, is given by
\begin{equation}
GM_{U}(r)=\frac{C}{\left( 1-n\right) r^{n-1}}+\frac{\left(
1+v_{tg}^{2}\right) v_{tg}^{2}}{1+v_{tg}^{2}+v_{tg}^{4}}r.
\end{equation}

Hence, the problem of the determination of the metric and of the
dark radiation and dark pressure in the constant tangential
velocity region around galaxies in the brane world model is
completely solved. It remains to determine the values of the
integrations constants $C$ and $C_{2}$.

To do this we assume that in the presence of matter the bulk
effects on the brane are negligible small. This definitely
represents a very good approximation for the galaxy, which has as
only constituent ``normal'' matter, with a density much higher than
the density of the matter outside the galaxy. On the other hand,
at large distances from the galactic center the cosmological
background dominates the dynamics.

Therefore, by assuming that the radius of the baryonic matter
distribution in the galaxy is $R$, we can match our solution for
$r=R$ with the Schwarzschild-de Sitter metric, so that
\begin{equation}
e^{\nu }{|}_{r=R}\approx 1-\frac{2GM}{R}-\frac{\Lambda }{3}R^{2},
\end{equation}
and
\begin{equation}
e^{\lambda }{|}_{r=R}\approx \left(1-\frac{2GM}{R}-\frac{\Lambda
}{3}R^{2}\right)^{-1},
\end{equation}
respectively. This gives
\begin{equation}
C_{2}=R^{-2v_{tg}^{2}}\left( 1-\frac{2GM}{R}\right) ,
\end{equation}
and
\begin{equation}
C=-\left( 1-n\right) \frac{\left( 1+v_{tg}^{2}\right) v_{tg}^{2}}{%
1+v_{tg}^{2}+v_{tg}^{4}}R^{n},
\end{equation}
respectively. Therefore the dark mass, the dark radiation and the dark
pressure terms can finally be written as
\begin{equation}
GM_{U}(r)=\frac{\left( 1+v_{tg}^{2}\right) v_{tg}^{2}}{1+v_{tg}^{2}+v_{tg}^{4}%
}r\left[ 1-\left( \frac{R}{r}\right) ^{n}\right] ,  \label{DM4}
\end{equation}
\begin{equation}
\alpha U(r)=\frac{1}{r^{2}}\left\{ \frac{\left( 1+v_{tg}^{2}\right)
v_{tg}^{2}}{3\left( 1+v_{tg}^{2}+v_{tg}^{4}\right) }\left[ 1-\left(
1-n\right) \left( \frac{R}{r}\right) ^{n}\right] -\Lambda r^{2}\right\} ,
\end{equation}
\begin{equation}
\alpha P(r)=-\frac{1}{r^{2}}\left[ \frac{\left( 1-n\right) \left(
1+v_{tg}^{2}\right) \left( 3+v_{tg}^{2}\right) }{3\left(
1+v_{tg}^{2}+v_{tg}^{4}\right) }\left( \frac{R}{r}\right) ^{n}-\Lambda r^{2}%
\right].
\end{equation}

Hence, the components of the Weyl tensor from the bulk can be
obtained in terms of the observable quantities, like the baryonic
radius of the galaxy and the tangential velocity of test particles
gravitating in stable circular orbits around the galactic center.

The brane world metric in the constant tangential velocity region
has a singularity at some $r=r_h$, which can be obtained by
solving the non-linear algebraic equation $\exp({-\lambda})=0$.
Depending on the numerical values of $n$ this equation may have
one or more positive roots, which define an event horizon.
Therefore the singularity at $r=0$ is hidden and cannot be seen by
an external observer.

\section{Qualitative analysis of the structure equations of the vacuum on
the brane for given simple equations of state of the dark pressure}
\label{seciv}

In the present Section we investigate the general behavior of the
tangential velocity on the brane by assuming a simple linear
equation of state for the dark pressure. Since generally the
structure equations of the vacuum on the brane cannot be solved
exactly, we shall analyze the dynamics of test particles by using
methods from the qualitative analysis of dynamical systems, by
closely following the approach of~\cite{boyce}.

We consider the case in which the dark pressure is proportional to the dark
radiation $P=\gamma U$, where $\gamma $ is an arbitrary constant, which can
take both positive and negative values. In the reduced variables $\mu $ and $%
p$ the linear equation of state is $p=\gamma \mu $, and the structure
equations of the gravitational field on the brane have the form
\begin{equation}
\frac{dq}{d\theta }=\mu -q,  \label{aut3a}
\end{equation}
\begin{equation}
\left( 1+2\gamma \right) \frac{d\mu }{d\theta }=2\left( 1-\gamma \right) \mu
-\frac{\left( \gamma +2\right) \mu \left[ q+\frac{1+2\gamma }{3}\mu \right]
}{1-q}.  \label{aut4a}
\end{equation}

The tangential velocity of the particles in stable circular orbit
is given by
\begin{equation}
v_{tg}^{2}=\frac{1}{2}\frac{q+\frac{1+2\gamma }{3}\mu }{1-q}.
\end{equation}

Let us firstly analyze the special case where $\gamma =-1/2$.
Then, by virtue if the second Eq.~(\ref{aut4a}), we obtain two
possible solution, either $q=\mu =0$ or $q=\mu =2/3$. Since both
solution are physically uninteresting, implying either $v_{tg}=0$
or $v_{tg}=1$, let us assume henceforth that $\gamma \neq -1/2$,
and rewrite the system of equation into the following form
\begin{equation}
\frac{dq}{d\theta }=-q+\mu , \label{aut3b}
\end{equation}
\begin{equation}
\frac{d\mu }{d\theta }=\frac{2(1-\gamma )}{1+2\gamma }\mu -\frac{\gamma +2}{%
1+2\gamma }\frac{\mu \left[ q+\frac{1+2\gamma }{3}\mu \right] }{1-q},
\label{aut4b}
\end{equation}
which is finally written as
\begin{equation}
\frac{d\xi }{d\theta }=A\xi +n,  \label{autuse}
\end{equation}
where we have denoted
\begin{equation}
\xi =
\begin{pmatrix}
q \\
\mu
\end{pmatrix}
,\quad A=
\begin{pmatrix}
-1 & 1 \\
0 & 2(1-\gamma )/(1+2\gamma )
\end{pmatrix}
,\quad n=
\begin{pmatrix}
0 \\
\frac{\gamma +2}{1+2\gamma }\frac{\mu \left[ q+\frac{1+2\gamma }{3}\mu
\right] }{1-q}
\end{pmatrix}
\end{equation}

The system of equations~(\ref{autuse}) has two critical points
$X_{0}=(0,0)$ and $X_{\gamma }=(3(1-\gamma )/(\gamma ^{2}+\gamma
+7),3(1-\gamma )/(\gamma ^{2}+\gamma +7))$. For $\gamma =1$, the
two critical points of the system coincide. Depending of the
values of $\gamma $, these points lie in different regions of the
phase space plane~$(q,\mu)$.

Since the term $||n|| / ||\xi|| \rightarrow 0$ as $||\xi|| \rightarrow 0$,
the system of equations~(\ref{autuse}) can be linearized at the critical
point $X_0$. The two eigenvalues of the matrix $A$ are given by $r_1=-1$ and
$r_2=2(1-\gamma)/(1+2\gamma)$, and determine the characteristics of the
critical point $X_0$. For $\gamma \in (-\infty,-1/2) \cup [1,\infty)$ both
eigenvalues are negative and unequal. Therefore, for such values of $\gamma$
the point $X_0$ is an improper asymptotically stable node.

If $\gamma \in (-1/2,1)$, we find one positive and one negative
eigenvalue, which corresponds to an unstable saddle point at the
point $X_0$. Moreover, since the matrix $dA/d\xi(X_0)$ has real
non-vanishing eigenvalues, the point $X_0$ is hyperbolic. This
implies that the properties of the linearized system are also
valid for the full non-linear system near the point $X_0$. It
should be mentioned however, that this first critical point is the
less interesting one from a physical point of view, since it
corresponds to the `trivial' case where both physical variables
vanish.

Before performing a similar analysis for the second critical point, it is
worth mentioning that the structure equations can be solved exactly for the value $%
\gamma=-2$. In that case, the non-linear term $n$ in
Eq.~(\ref{autuse}) identically vanishes, and the system of
equations becomes a simple linear system of differential
equations. For $\gamma=-2$ the two eigenvalues of $A$ are given by
$r_1=-1$ and $r_2=-2$, the two corresponding eigenvectors are
linearly independent and the general solution can be written as
follows
\begin{equation}
\xi_{\gamma=-2} = (q_0 + \mu_0)
\begin{pmatrix}
1 \\
0
\end{pmatrix}
e^{-\theta} + \mu_0
\begin{pmatrix}
-1 \\
1
\end{pmatrix}
e^{-2\theta},
\end{equation}
where $q_0 = q(0)$ and $\mu_0 = \mu(0)$. One can easily transform this
solution back into the radial coordinate $r$ form by using $\theta=\log(r)$
and get
\begin{equation}
\mu_{\gamma=-2} = \frac{\mu_0}{r^2},\quad q_{\gamma=-2} = \frac{q_0}{r} +
\mu_0 \left(\frac{1}{r}-\frac{1}{r^2} \right).
\end{equation}

This solution can be used to analyze the resulting tangential
velocity profile as a function of the radius~$r$, which turn out
to have the same form as predicted by the Newtonian analysis.
Therefore such an equation of state, namely $p=-2\mu$, cannot
account for flat rotation curves in galaxies, see the end of
Section~\ref{secii}.

Let us now analyze the qualitative behavior of the second critical point $%
X_\gamma $. To do this, one has to Taylor expand the right-hand sides of Eqs.~(%
\ref{aut3b}) and~(\ref{aut4b}) around $X_\gamma$ and obtain the matrix $\tilde{A}$
which corresponds to the system, linearized around $X_\gamma$. This linearization
is again allowed since the resulting non-linear term, $\tilde{n}$ say, also
satisfies the condition $||\tilde{n}|| / ||\xi|| \rightarrow 0$ as $||\xi||
\rightarrow X_\gamma$. The resulting matrix reads
\begin{equation}
\tilde{A}=\begin{pmatrix}
-1 & 1 \\
\frac{3(-\gamma^2+5\gamma-4)}{(2+\gamma)^2 (1+2\gamma)} & \frac{\gamma -1}{2+\gamma}
\end{pmatrix}
\label{ew}
\end{equation}
and its two eigenvalues are given by
\begin{equation}
r_{\pm} = \frac{-3-6\gamma \pm \sqrt{16\gamma^4 + 8\gamma^3 + 132\gamma^2
-28\gamma -47}} {4\gamma^2+10\gamma+4}.
\end{equation}

For $-0.5 < \gamma < 0.674865$ the argument of the square root
becomes negative and the eigenvalues complex. Moreover, the values
$\gamma=-1/2$ and $\gamma=-2$ have to be excluded, since the
eigenvalues in Eq.~(\ref{ew}) are not defined in these cases.
However, both cases have been treated separately above.

\begin{figure}[!ht]
\centering
\includegraphics[width=.48\textwidth]{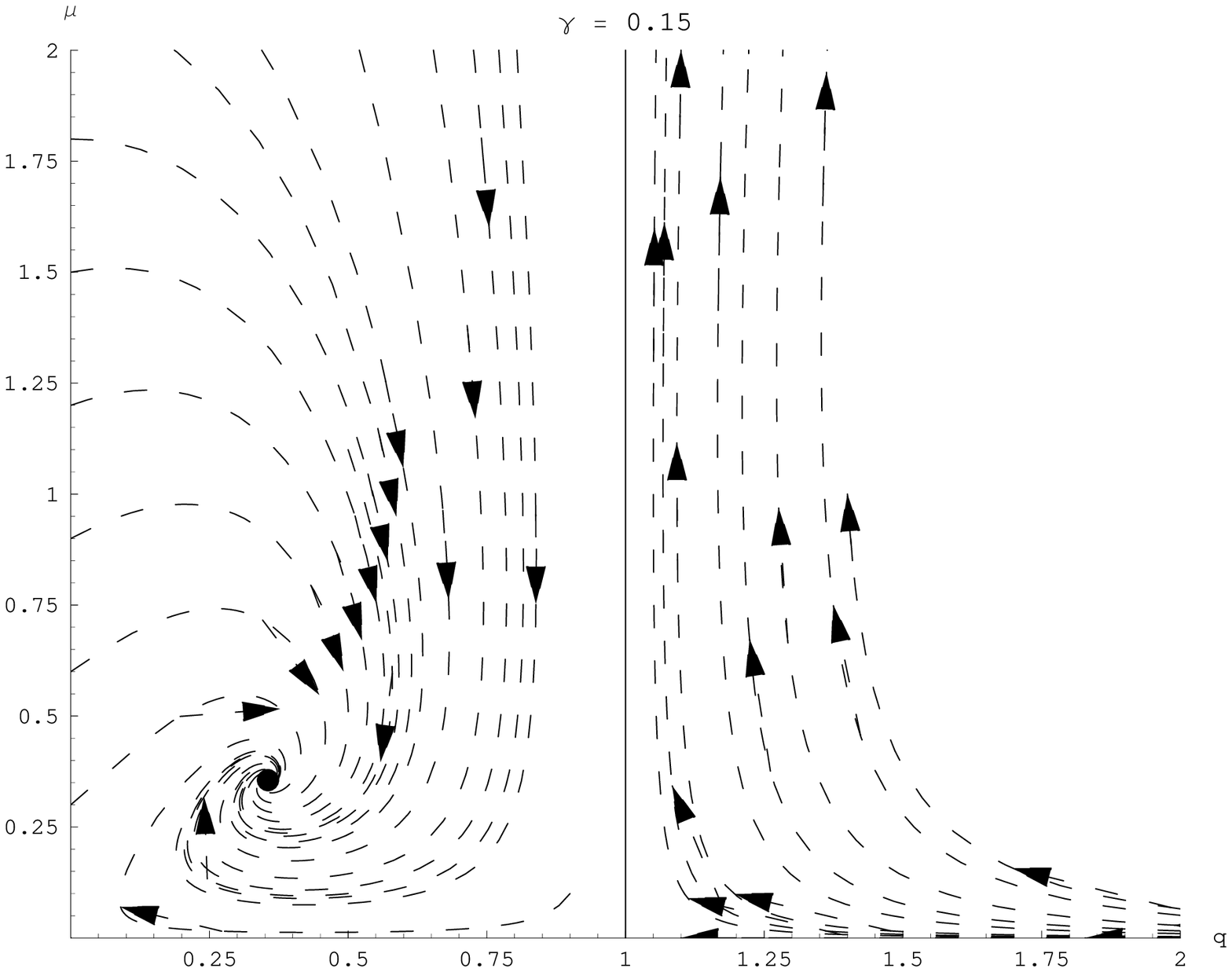}
\includegraphics[width=.48\textwidth]{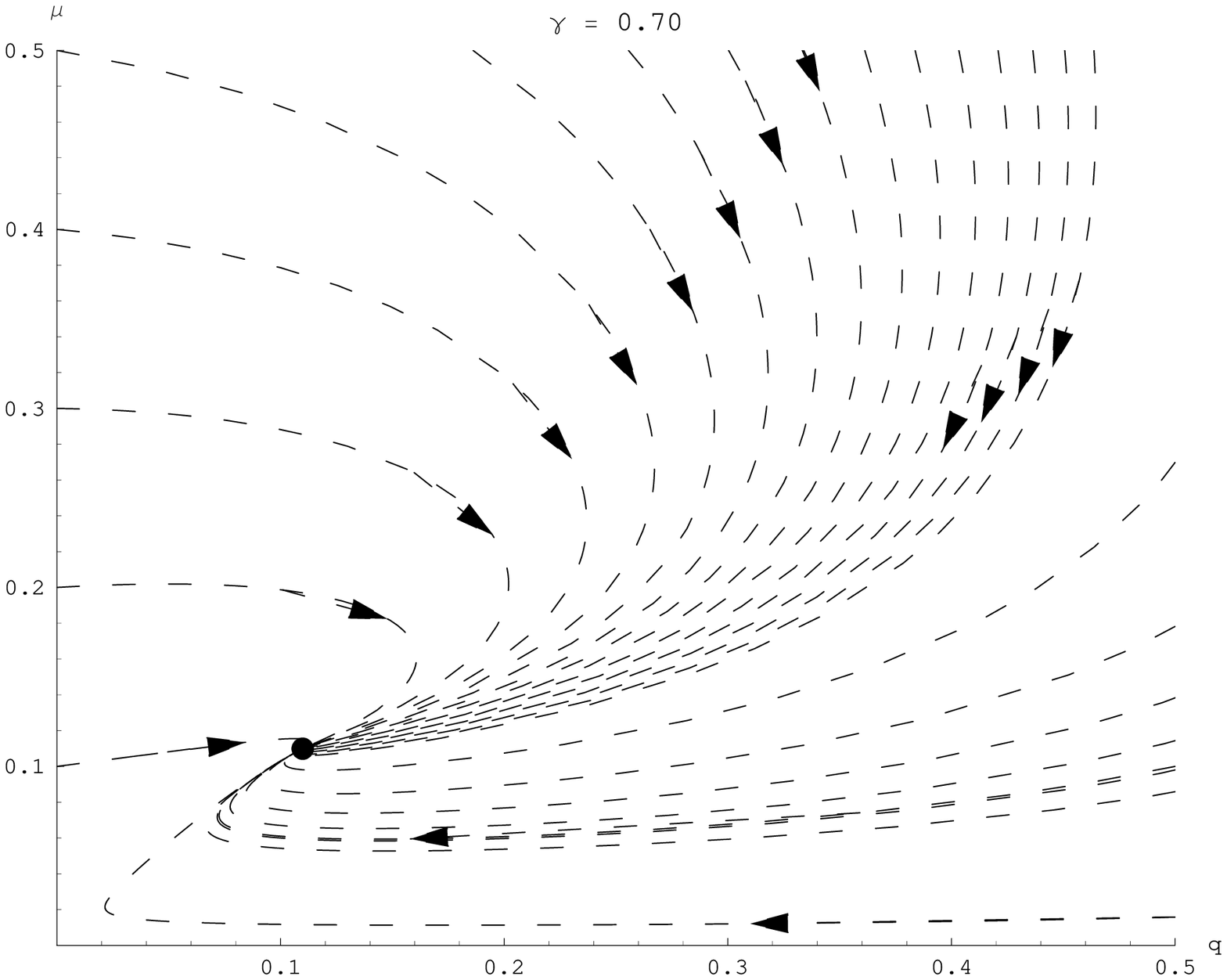}
\caption{The left figure shows the phase space plot $\mu=\mu(q)$ of the system
         (\ref{aut3b})--(\ref{aut4b}) for $\gamma=0.15$. The dot
         represents the critical point $X_\gamma $ and one clearly
         sees that $X_\gamma $ corresponds to an asymptotically stable
         spiral point. In the right figure $\gamma=0.70$, and there
         the critical point $X_\gamma $ corresponds to an asymptotically
         stable improper node.}
\label{fig1}
\end{figure}

If $\gamma \in (-\infty ,-1/2)\cup (1,\infty )$, then $X_{\gamma}$
corresponds to an unstable saddle point and for $\gamma \in
(0.674865,1)$ it corresponds to an asymptotically stable improper
node. More interesting is the parameter range $\gamma \in
(-1/2,0.674865)$ where the eigenvalues become complex, however,
their real parts are negative definite. Hence, for those values we
find an asymptotically stable spiral point at $X_{\gamma}$. Since this
point is also an hyperbolic point, the described properties are
also valid for the non-linear system near that point, see
Fig.~\ref{fig1}. This qualitative study of the behavior of the
structure equations of the vacuum on the brane shows that not all
equations of state of the dark pressure would predict flat
rotation curves.

With Eq.~(\ref{eos}) we were able to explicitly construct the
equation of state that yields rotation curves of constant
velocity. Since these velocities are about $200-300 {\rm km/s}$,
we obtain as an equation of state
\begin{equation}
p\approx (3-7)\times 10^6 \mu -3 ,
\end{equation}
which is of the general form $p=a\mu-b$.

The above qualitative analysis can easily be repeated for the
physically relevant equation of state. The first difference is
that $X_0=(0,0)$ is no longer a critical point to the system, and
the critical point $X_\gamma$ gets a contribution due to the
parameter $b$. Moreover, $X_\gamma$ is replaced by a pair of
critical points, related to the two square roots of a quadratic
equation that reduces to a linear equations as $a\rightarrow 0$.
The critical point in the $p>0$ and $\mu>0$ region in phase space
corresponds to an asymptotically stable improper node, and the
qualitative picture is that of the right plot of Fig.~\ref{fig1}.

\section{Discussions and final remarks}
\label{secv}

The galactic rotation curves and the mass distributions in clusters of
galaxies continue to pose a challenge to present day physics. One would like
to have a better understanding of some of the intriguing phenomena
associated with them, like their universality and the very good correlation
between the amount of dark matter and the luminous matter in the galaxy. To
explain these observations, the most commonly considered models are based on
particle physics in the framework of Newtonian gravity, or of some
extensions of general relativity~\cite{expl}.

In the present paper we have considered, and further developed, an
alternative view to the dark matter problem~\cite{Ma04,Ha05},
namely, the possibility that the galactic rotation curves and the
mass discrepancy in clusters of galaxies can naturally be
explained in models in which our Universe is a brane in a
multi-dimensional spacetime. The extra terms in the gravitational
field equations on the brane induce a supplementary gravitational
interaction, which can account for the observed behavior of the
galactic rotation curves. As one can see from Eq.~(\ref{DM4}), in
the limit of large $r \gg R$, and by taking into account that
$v_{tg}^{2} \ll 1$, we obtain $GM_{U}\approx v_{tg}^{2}r$, a
behavior which is perfectly consistent with the observational
data~\cite{Bi87}, and is usually attributed to the existence of
the dark matter. Since inside the galaxy $M_U\approx 0$, it
follows that the constant tangential velocity of the test
particles is determined by the baryonic mass $M_B$ and the radius
$R$ of the galaxy by the equation $v_{tg}^2\approx GM_B/R$. This
gives for the dark mass the scaling relation
\begin{equation}
\frac{M_U}{M_B} \approx \frac{r}{R}.
\end{equation}

By using the simple observational fact of the constancy of the
galactic rotation curves, the galactic metric and the
corresponding components of the projected Weyl tensor $E_{\mu \nu
}$ (dark radiation and dark pressure) can be completely
reconstructed, without any supplementary assumption.

Weak lensing provides a unique probe of the gravitational
potential on large scales. Hence, in the context of dark matter,
it can provide constrains on the extent and shapes of dark matter
halos. Furthermore, it can test alternative theories of gravity
(without dark matter). Due to the fixed form of the galactic
metric on the brane, in our model the light bending angle is a
function of the tangential velocity of particles in stable
circular orbit and of the baryonic mass and radius of the galaxy.
The specific form of the bending angle is determined by the brane
galactic metric, and this form of the metric is different as
compared to the other dark matter models (long-range
self-interacting scalar fields, MOND, non-symmetric gravity etc.).
As shown in \cite{Pal,Majum05,Wh05,Ha05}, the gravitational light
deflection angle is much larger than the value predicted by the
standard general relativistic approach. Even when we compare our
results with standard dark matter models, like the isothermal dark
matter halo model, we still may find significant differences in
the lensing effect. Therefore the study of the gravitational
lensing may provide the evidence for the existence (or
non-existence) of the bulk effects on the brane. On the other
hand, since in this model there is only baryonic matter, all the
physical properties at the galactic level are determined by the
amount of the luminous matter and its distribution.

To test alternative theories of gravity by using lensing one can
use two approaches \cite{Hoek02}. A measurement of the radial
dependence of the lensing signal gives the best accuracy. This
method requires the knowledge of the deflection law. The most
direct test is the detection of the azimuthal variation of the
lensing signal around the lenses, and the measurements does not
require knowledge of the deflection law. Any anisotropy in the
lensing signal caused by the galaxy decreases as $r^{-2}$ and
therefore is negligible in the bulk effects dominated region.

The detection of an anisotropy in the lensing signal around
galaxies may provide the best test for discriminating between dark
matter and alternative gravity models, including the present
model. The results of a study of weak lensing by galaxies based on
45.5 deg$^2$ of $R_C$ band imaging data from the Red-Sequence
Cluster Survey were presented in \cite{Hoek04}. A significant
flattening was detected, which implies that the halos are well
aligned with the light distribution, and an isotropic lensing
signal is excluded with 99.5\% confidence.

From an observational point of view an important problem is to
estimate an upper bound for the cutoff of the constancy of the
tangential velocities. The idea is to consider the point at which
the decaying density profile of the dark radiation associated to
the galaxy becomes smaller than the average energy density of the
Universe, and the cosmological expansion dominates the dynamics of
the particles. Let the value of the coordinate radius at the point
where the two densities are equal be $R_{U}$. Then at this point
$3\alpha\,U(R_{U}) =(8\pi G) \rho_{univ}$, where $\rho_{univ}$ is
the mean energy density of the universe.

An alternative estimation of $R_{U}$ can be obtained from the
observational requirement that at the cosmological level the
energy density of the dark matter (which in our case is the dark
mass associated to the dark radiation) represents a fraction
$\Omega _{DM}\approx 0.25$ of the total energy density of the flat
universe, with density parameter $\Omega =1$. Therefore the ``dark
matter'' contribution inside a ball of radius $R_{U}$ is given by
$4\pi \Omega_{DM}R_{U}^{3}\rho_{crit}/3$, which gives
\begin{equation}
R_{U}\approx \frac{1}{\sqrt{2\Omega _{DM}}}\frac{1}{h(z)H_{0}}v_{tg},
\end{equation}
where $h(z)$ is the Hubble constant normalized to its local value: $%
h^{2}(z)=\Omega _{m}\left( 1+z\right) ^{3}+\Omega _{\Lambda }$, $%
\Omega _{m}$ is the total mass density parameter and $\Omega _{\Lambda }$ is
the dark energy density parameter. For a tangential velocity of the order of
200 km/s it follows that $R_{U}\approx 2.8$ Mpc. Therefore, we predict
that the flat rotation curves region should extend up to distances of a few
megaparsecs from the galactic center.

If we assume that the flat rotation curves extend indefinitely,
the resulting spacetime is not asymptotically flat, but of de
Sitter type. This is due to the presence of the cosmological
constant $\Lambda $ on the brane. Observationally, the galactic
rotation curves remain flat to the farthest distances that can be
observed. Therefore, from both the observational and the
theoretical point of view it is important to estimate the possible
role of the cosmological constant on the extra galactic dynamic.
By assuming that the cosmological constant has a numerical value
of the order of $\Lambda \approx 3\times 10^{-56}$ cm$^{-2}$
\cite{PeRa03}, at a distance of $r=100$ kpc, for a galaxy with
mass $M=10^{10}M_{\odot}$, the quantities $GM/r\approx 4.94\times
10^{-9}$ and $\Lambda r^2\approx 2.7\times 10^{-9}$ are roughly of
the same order of magnitude. For clusters of galaxies with masses
of the order of $10^{14}M_{\odot }$ and radii of the order of $2$
Mpc, we have $GM/r\approx 2.47\times 10^{-6}$ and $\Lambda
r^2\approx 1.08\times 10^{-6}$, respectively.

Hence the presence of a non-zero cosmological constant may play an
important role in the dynamics of the test particles in stable
circular orbits at the outer boundary of the galaxies or clusters
of galaxies, since it may generate physical effects that are of
the same order of magnitude as the extra-dimensional effects from
the bulk. Therefore, for the correct estimation in regions far
away from the galactic center of the dark radiation, of the dark
pressure, of the tangential velocities of test particles and of
other related effects (like, for example, the bending angle of the
light) one must take into account the effect of the cosmological
constant on the brane.

In the present model all the relevant physical quantities,
including the dark mass (playing the role of the dark matter on
the brane), the dark radiation and the dark pressure describing
the non-local effects due to the gravitational field of the bulk,
are expressed in terms of observable parameters -- the tangential
velocity, the baryonic mass and the radius of the galaxy.

There are three observationally testable predictions of the model:
a) the constant rotation curve region may extend much farther than
presently observed, up to a few megaparsecs, b) the weak lensing
effect in the dark radiation dominated region is different from
that predicted by the standard dark matter models, and other
alternative gravitational theories and c) the numerical value of
the maximum rotational velocity is mainly determined by the
``normal'' matter content of the galaxy.

As for the main advantages of our model, on the one hand it is
esthetically attractive -- ``matter'' is a manifestation of
geometry, it is relatively simple and, on the other hand, it is
directly testable observationally. The best testing ground for
brane world models is the extra-galactic space, where the effects
of galactic/extra-galactic matter perturbations are minimal, and
where we expect that the bulk effects appear in a "pure" form, and
they dominate the dynamics of test particles (gas clouds).
Moreover, the extra-galactic space is accessible to direct optical
or radio observation. In the solar system the bulk effects are
very small, and difficult to observe.

Therefore, these results open the possibility of testing the brane
world models by using direct astronomical and astrophysical
observations at the galactic or extra galactic scale. In this
paper we have provided some basic theoretical tools necessary for
the in depth comparison of the predictions of brane world models
and of the observational results.

\acknowledgments

The authors would like to thank to the two anonymous referees for
suggestions that helped us to significantly improve the
manuscript, and to Roy Maartens for his valuable comments. The
work of CGB was supported by research grant BO 2530/1-1 of
the German Research Foundation (DFG). The work of TH was
supported by the RGC grant No.~7027/06P of the government of the
Hong Kong SAR.

\end{document}